\documentclass[12pt]{article}
\usepackage{wider}
\begin{document}

\newcommand{\sqvb}{\ensuremath{ \langle \!\langle 0 |} }
\newcommand{\sqvk}{\ensuremath{ | 0 \rangle \!\rangle } }
\newcommand{\sqvn}{\ensuremath{ \langle \! \langle 0 |  0 \rangle \! \rangle} }
\newcommand{\wh}{\ensuremath{\widehat}}
\newcommand{\be}{\begin{equation}}
\newcommand{\ee}{\end{equation}}
\newcommand{\bea}{\begin{eqnarray}}
\newcommand{\eea}{\end{eqnarray}}
\newcommand{\ra}{\ensuremath{\rangle}}
\newcommand{\la}{\ensuremath{\langle}}
\newcommand{\rra}{\ensuremath{ \rangle \! \rangle }}
\newcommand{\lla}{\ensuremath{ \langle \! \langle }}
\newcommand{\str}{\rule[-.125cm]{0cm}{.5cm}}
\newcommand{\pr}{\ensuremath{^\prime}}
\newcommand{\ppr}{\ensuremath{^{\prime \prime}}}
\newcommand{\da}{\ensuremath{^\dag}}
\newcommand{\as}{^\ast}
\newcommand{\eps}{\ensuremath{\epsilon}}
\newcommand{\ve}{\ensuremath{\vec}}
\newcommand{\ka}{\kappa}
\newcommand{\non}{\ensuremath{\nonumber}}
\newcommand{\lf}{\ensuremath{\left}}
\newcommand{\rt}{\ensuremath{\right}}
\newcommand{\al}{\ensuremath{\alpha}}
\newcommand{\dfn}{\ensuremath{\equiv}}
\newcommand{\ga}{\ensuremath{\gamma}}
\newcommand{\ti}{\ensuremath{\tilde}}
\newcommand{\wti}{\ensuremath{\widetilde}}
\newcommand{\hs}{\ensuremath{\hspace*{.5cm}}}
\newcommand{\bet}{\ensuremath{\beta}}

\newcommand{\cO}{\ensuremath{{\cal O}}}
\newcommand{\cS}{\ensuremath{{\cal S}}}
\newcommand{\cF}{\ensuremath{{\cal F}}}

\newcommand{\pup}{\ensuremath{^{(p)}}}

\title{\bf Relative Frequency and Probability in the Everett Interpretation of Heisenberg-Picture Quantum Mechanics\thanks{
This work was sponsored by the Air Force under Air Force Contract 
F19628-00-C-0002.  Opinions, interpretations, conclusions, and 
recommendations are those of the author and are not necessarily endorsed 
by the U.S. Government.
}
}
\author{
Mark A. Rubin\\
\mbox{}\\   
Lincoln Laboratory\\ 
Massachusetts Institute of Technology\\  
244 Wood Street\\                         
Lexington, Massachusetts 02420-9185\\      
rubin@LL.mit.edu\\ 
}
\date{\mbox{}}
\maketitle


\begin{abstract}

The existence of probability in the sense of the frequency interpretation, i.e.   
probability as ``long term relative frequency,'' is shown to
follow from the dynamics and the interpretational rules of
Everett quantum mechanics in the Heisenberg picture.
This proof is free of the  difficulties encountered in applying
to the Everett interpretation
previous results regarding relative frequency and probability in quantum mechanics.
The ontology of the Everett interpretation  in the Heisenberg picture is also discussed.

\noindent Key words: relative frequency, probability, Everett interpretation, Heisenberg
picture, Born rule
\end{abstract}

\section{Introduction} \label{sec_intro}

The Everett interpretation of quantum mechanics \cite{Everett57} posits that
all physical phenomena can be described by unitary time evolution.
In the standard Copenhagen interpretation of quantum mechanics (see, e.g.,
\cite{dEspagnat76}), the phenomenon
of {\em probability}\/ arises by virtue of ``reduction of the wavefunction,''
an explicitly {\em non}\/unitary type of
time evolution   which, when measurements
are made, supplants
the unitary evolution that otherwise occurs. A challenge for the
Everett interpretation is, therefore, to show that it {\em predicts}\/
the existence of probability  in the context of  completely unitary  time evolution.

Probability, whether 
in  a quantum-mechanical or classical context, has two characteristic aspects\cite{Cramer46}. One of these is
{\em randomness}\/, 
the fact that the result of one repetition of an experiment cannot in general be
predicted with certainty.  Everett quantum mechanics predicts and explains
the existence of  randomness in a straightforward manner. Suppose an experimenter, Alice,
performs a quantum experiment to measure an observable with two possible outcomes, ``up'' and
``down.'' After Alice has performed the experiment and measured the observable,
there exist two noninteracting copies of Alice, each having the same memories
as the other except with regard to the result of the experiment---we may
call them ``Alice-who-saw-up'' and ``Alice-who-saw-down.'' To the question of
which Alice 
is  the ``real Alice,''  the unequivocal answer is, ``both are.'' Thus, in the
Everett interpretation,  the question ``which 
result will Alice see'' 
does not even in principle have an answer. (See discussion and references in
\cite{Vaidman02}).

The other characteristic aspect of probability referred to above is the {\em regularity}\/ of 
the relative frequency of the results of an experiment when it is
repeated a large number of times  or, equivalently, when a large ensemble
of identical experiments are performed.
``Whenever we say that the probability of an event $E$\/ with respect to an experiment
${\cal E}$\/ is equal to $P$\/, the concrete meaning of this assertion [is] the following: In a long series of repetitions of ${\cal E}\/$, it is practically
certain that the [relative] frequency of $E$\/ will be approximately equal to $P$\/
\ldots This statement will be referred to as the frequency interpretation of the 
probability $P$\/ \cite[pp. 148-9]{Cramer46}.'' Or, more succinctly, probability is
``the theoretical value of long range relative frequency \cite[p. 58]{Polya54}.''

These two properties are equally true of probability whether the setting is quantum
or classical.  In the  case of quantum-mechanical probability, there is
another key property: namely, the Born rule for the value of the probability.  
The possible results of a measurement corresponding to an operator $\wh{a}$\/
are the eigenvalues $\al_i$\/ of $\wh{a}$\/, 
\begin{equation}
\wh{a}|\al_i\ra=\al_i|\al_i\ra, \hspace*{.5cm} i=1,2, \ldots,n. \label{eigenvalue_eq}
\end{equation}
The probability $P(\al_i)$\/ of measuring the result $\al_i$\/ when, immediately before the measurement,
the system
being measured is  in the state
\begin{equation}
|\psi_0\ra =c_1 |\al_1\ra +  \ldots + c_n|\al_n\ra, \label{superposition}
\end{equation}
where the $c_i$\/'s are complex numbers satisfying
\begin{equation}
\sum_{i=1}^n |c_i|^2=1, \label{normalization_condition}
\end{equation}
is given by the Born rule
\begin{equation}
P(\al_i)= 
|c_i|^2. \label{Born_rule}
\end{equation}
(We assume here and in the remainder of this paper that the
$\al_i$\/ are nondegenerate.)

A proof has been presented by 
Hartle\cite{Hartle68} 
that probability,
as characterized by the above-mentioned properties, exists in unitary quantum mechanics
without having to be incorporated  via  wavefunction reduction.
(See also \cite{Finkelstein63}.)
This proof makes use of the interpretive rule of quantum mechanics which 
states that if a system is in an eigenstate corresponding the eigenvalue
$\al_i$\/ of the operator $\wh{a}$\/, then a measurement of the observable 
$\wh{a}$\/ on the system will with certainty yield the result $\al_i$\/.
Hartle makes use of   operators $\wh{f}_N(\al_i),$\/ $i=1, \ldots, n$\/,
corresponding to the measurement of the relative frequency of the result $\al_i$\/ in 
an ensemble of $N$\/ identical systems (i.e., the fraction of  systems
in the ensemble in which  the result $\al_i$\/ is obtained upon measurement)
\cite{Hartle68,Graham73}, 
and shows that, in the 
limit $N \rightarrow \infty$\/,
the state vector of the entire ensemble
\begin{equation}
|\psi_N\ra=|\psi^{(1)}\ra |\psi^{(2)}\ra \ldots |\psi^{(N)}\ra \label{psi_sub_N}
\end{equation}
(where each $|\psi^{(p)}\ra$\/ is of the form (\ref{superposition}) with the same
$c_i$\/'s)
approaches an eigenstate of the relative frequency operators $\wh{f}_N(\al_i)$\/
with respective eigenvalues given by the Born rule (\ref{Born_rule}):
\begin{equation}
\lim _{N \rightarrow \infty }\wh{f}_N(\al_i)|\psi_N \ra=|c_i|^2 \lim _{N \rightarrow \infty }|\psi_N\ra. \label{lim_eigenvalue}
\end{equation}
Farhi, Goldstone and Gutmann\cite{FarhiGoldstoneGutmann89} prove a similar result
working from the outset in the $N=\infty$\/ Hilbert space 

As has been pointed out by Kent\cite{Kent90} and by Zurek\cite{Zurek98}, these proofs do
not demonstrate that the phenomenon of
probability exists in the Everett version of quantum mechanics.   
That  is because, in the Everett interpretation, the 
state vector (\ref{psi_sub_N})
is the state vector 
of the entire ``multiverse'' \cite{Deutsch96} and  describes all the Everett
branches at once. The state vector (\ref{psi_sub_N}), using (\ref{superposition}),
is equal to
\begin{equation}
|\psi_N\ra= \sum_{i^{(1)}=1}^n \ldots \sum_{i^{(N)}=1}^n 
c_1^{r_1}  \ldots c_n^{r_n} |i^{(1)},  \ldots , i^{(N)}\ra,   \label{sum_of_branches}
\end{equation}
where
\begin{equation}
r_i=r_i(i^{(1)}, \ldots,i^{(N)})=\sum_{p=1}^n \delta_{i,i^{(p)}}
\end{equation}
is the number of factors of eigenvalue-$i$\/ states in a given term in 
(\ref{sum_of_branches}),  and where each of the $n^N$\/  states
\begin{equation}
c_1^{r_1}  \ldots c_n^{r_n}|i^{(1)},  \ldots , i^{(N)}\ra = 
c_1^{r_1}  \ldots c_n^{r_n}|\al_{i^{(1)}} \ra \ldots |\al_{i^{(N)}}\ra, \label{branch}
\end{equation}
is termed a 
``branch'' of the overall state vector (\ref{sum_of_branches}).
In the Everett interpretation,  as usually formulated in terms of
state vectors, each branch (\ref{branch})
is deemed to correspond to a distinct physical reality.
If an observer $\cO$\/ measures each of the $N$\/ systems
in the ensemble, the state vector describing the observer and the
ensemble is, after the measurement has taken place,
\begin{equation}
|\cO, \psi_N\ra= 
\sum_{i^{(1)}=1}^n \ldots \sum_{i^{(N)}=1}^N 
c_1^{r_1}\ldots c_n^{r_n} 
|\cO; i^{(1)},  \ldots , i^{(N)}\ra  |i^{(1)}, \ldots , i^{(N)}\ra.   \label{sum_of_branches_O}
\end{equation}
where the factor
\begin{equation}
|\cO; i^{(1)},   \ldots , i^{(N)}\ra
\end{equation}
in each branch
\begin{equation}
c_1^{r_1}  \ldots c_n^{r_n}|\cO; i^{(1)},  \ldots , i^{(N)}\ra  |i^{(1)},  \ldots , i^{(N)}\ra \label{branch_O}
\end{equation}
corresponds to a 
distinct perception on the part of the observer, correlated with
definite measurement results for each of the systems in
the ensemble. 

What must be shown to establish the frequency interpretation of probability and the Born rule
in Everett quantum mechanics is that an observer performing a relative frequency
measurement on an ensemble of independent identical systems, in the limiting
case that the number of systems making up the ensemble becomes infinite, will 
always (never) have the
perception of measuring a relative frequency equal to (different from) the Born-rule
probability (\ref{Born_rule}). In terms of the branches (\ref{branch_O}),
what must be shown is that, in the limit $N \rightarrow \infty$\/,
the observer in each branch perceives the Born-rule relative frequency.
The complete state vector (\ref{sum_of_branches_O})
does not correspond to an Everett branch in which an observer has experienced specific 
measurement results for each of the systems in the ensemble---else it
would be of the form (\ref{branch_O})---so demonstrating that 
the complete state vector is an eigenfunction of the relative frequency
operators $\wh{f}_N(\al_i)$\/ with  the corresponding Born-rule  eigenvalues
(\ref{Born_rule})
does not make the
case that must be made.

DeWitt\cite{DeWitt72} and Okhuwa\cite{Okhuwa93}
demonstrate that the state vector 
$|\chi_N\ra$\/ which is the sum of all of the
branches experiencing non-Born relative frequencies has zero norm in the  limit
of an infinitely-large ensemble,
\begin{equation}
\lim_{N \rightarrow \infty}\la \chi_N|\chi_N\ra=0. \label{zero_norm}
\end{equation}
Since experience inheres in individual branches, it is not relevant to
the task of characterizing which  experiences are or are not possible  
that some branches can 
or cannot be grouped
together so that the norm of their sum does or does not vanish.
(It is  true
that the norm of a branch corresponding to any particular outcome
of measurement of the entire ensemble will go to zero as the size of the ensemble, and hence
the number of possible outcomes, goes to infinity.  But this is as true
for branches in which Born-rule relative frequencies are experienced as
it is for branches in which relative frequencies differ from the Born rule 
(\ref{Born_rule}).)

Hartle and Farhi, Goldstone and Gutmann work in the Schr\"{o}dinger picture. DeWitt
and Okhuwa work in the Heisenberg picture, but focus on states rather than
operators. Recently, Deutsch and Hayden \cite{DeutschHayden00} and the author 
\cite{Rubin01,Rubin02}
have shown that the issue of locality in Everett quantum theory is 
clarified when one works in the Heisenberg picture and focuses on the dynamics of operators.
In this paper I show that such an approach clarifies as well
the analysis of the origin of probability.
In Sec. 2 below I  
review the formalism and interpretive rules of Heisenberg-picture 
Everett quantum mechanics, and introduce a rule which corresponds to
the interpretation given to  (\ref{zero_norm}) by  DeWitt and Okhuwa  
but which is free of the above-mentioned difficulties.  In Sec. 3 I show
that these rules, applied to a physically-realizable relative-frequency-measuring 
device---one of finite resolution, as described  by Graham\cite{Graham73}---lead to
the conclusion that the device will, to within the limits of its 
resolution, observe a result consistent with the Born rule (\ref{Born_rule})
in the limiting case of an infinitely-large ensemble.
Sec. {\ref{sec_summary} contains a summary
of the proof.
Sec. \ref{sec_discussion} discusses the ontology  of Everett 
quantum mechanics in the Heisenberg picture, and comments on some other frequency-based approaches to probability in the Everett interpretation.\footnote{ 
Approaches to  
probability in Everett or other
interpretations of quantum mechanics which  use 
concepts of probability other than the frequency interpretation, 
such as 
ignorance\cite{Vaidman02,Vaidman98}, decision 
theory\cite{Deutsch99,DeWitt98, Polley99,Polley01},
and Bayesian inference\cite{CavesFuchsSchack}, are outside the scope of this paper.}

\section{Measurement in Everett Quantum Mechanics} \label{sec_MEQM}

\subsection{Schr\"{o}dinger Picture} \label{meas_schrod_pic}

An observer $\cO$\/ measures an observable $\wh{g}$\/ pertaining to a  system $\cS$\/.
The eigenspectrum of $\wh{g}$\/ will be taken to finite, discrete and
nondegenerate:
\begin{equation}
\wh{g}|\cS;\ga_i \ra = \ga_i|\cS;\ga_i \ra, \hspace*{.5cm} i=1,\ldots, M.
\end{equation}
The state space of $\cO$\/ is spanned by $L+1$\/ eigenstates of the
operator $\wh{b}$\/ corresponding to the observer's state of 
belief regarding the result of the measurement:\footnote{``Observer'' 
here can refer as well to a machine as to a human being, with ``state of belief'' 
denoting results stored in a memory device.}
\begin{equation}
\wh{b}|\cO;\bet_i\ra= \bet_i |\cO;\bet_i\ra, \hs i=0,\ldots L,
\end{equation}
where $\bet_0$\/ corresponds to the state of ignorance prevailing before
a measurement has been made.
We do not assume that the measurement produces a one-to-one mapping from the
$\ga_i$\/'s to the $\bet_i$\/'s,  $i > 0$\/, but allow for the
possibility that several different $\ga_i$\/'s might correspond to the
same $\bet_i$\/:
\be 
\{\bet_i\}=\{\bet(\ga_j)\},\hs i=1,\ldots, L, \hs j=1,\ldots,M, \hs L \leq M,
\ee
\be
\bet_i \neq \bet_j, \hs i \neq j
\ee
with the values of $\bet_i$\/ otherwise arbitrary.

We will always consider situations in which
the state before the measurement has occurred, $|\psi (t_0)\ra $\/,
is of the form
\be
|\psi (t_0)\ra = |\cO;\bet_0\ra |\cS;\psi\ra; \label{uncorrelated_state}
\ee 
i.e., $\cO$\/ is in a state of ignorance and 
$\cS$\/ is an arbitrary state the observables of
which are uncorrelated with those of $\cO$\/.

If before the measurement $\cS$\/ is  in the
eigenstate of $\wh{\gamma}$\/ with eigenvalue $\gamma_i$\/, then
the state vector describing both $\cO$\/ and $\cS$\/ before the measurement
is the product state
\begin{equation}
|\cO; \bet_0\ra |\cS;\ga_i\ra. \label{start_state_i}
\end{equation}
By virtue of the interpretation of operator eigenstates as states in which measurement
will definitely obtain the corresponding eigenvalue as a result, 
the action
on the initial state (\ref{start_state_i}) of the unitary operator $\wh{U}$\/ corresponding to  $\cO$\/ measuring
$\cS$\/ must be to produce the  state in which $\wh{g}$\/
definitely has the value $\ga_i$\/ and $\wh{b}$\/ definitely has the value
$\bet(\ga_i)$\/:
\be
\wh{U}|\cO; \bet_0\ra |\cS; \ga_i\ra=
|\cO; \bet(\ga_i)\ra |\cS; \ga_i\ra.\label{U_action_i}
\ee
From this result and the linearity of quantum mechanics,  we conclude that
\be
\wh{U}=\sum_{i=1}^L \wh{u}_i \otimes \wh{P}_i^{\wti{\cS}}  \label{U}
\ee
where $\wh{P}_i^{\wti{\cS}}$\/ is the projection operator 
which projects out those states corresponding
to $\bet_i$\/:
\be
\wh{P}_i^{\wti{\cS}}=\sum_{j|\bet(\ga_j)=\bet_i}\wh{P}_j^{\cS},\hs i=1, \ldots, L,
	\label{P_S_tilde}
\ee
\be
\wh{P}_j^{\cS}=|\cS;\ga_j \ra \la \cS;\ga_j |, \hs j=1, \ldots, M.
	\label{P_S}
\ee
The action of $\wh{u}_i$\/ in the state space of $\cO$\/ is 
\be
\wh{u}_i |\cO;\bet_0\ra =|\cO;\bet_i\ra, \hs i=1,\ldots, L. \label{u_i_action}
\ee
The action of $\wh{u}_i$\/ on the other states of $\cO$\/,
$|\cO;\bet_i\ra$\/ with $i >0$\/, will not play a role in subsequent
analysis. (For an example of the complete action of $\wh{u}_i$\/  in
the state space of an observer with $L=2$\/ see \cite[Sec. 4.1]{Rubin01}.)

\subsection{Heisenberg Picture} \label{subsec_HP}

We will take the initial-time constant Heisenberg-picture state vector
to be the before-measurement state (\ref{uncorrelated_state}),
and we will distinguish Heisenberg-picture operators by
explicit time arguments.  At the initial time, $t_0$\/,
the Heisenberg-picture operators are equal to the corresponding Schr\"{o}dinger-picture
operators:
\be
\wh{g}(t_0)=\wh{g},
\ee
\be
\wh{b}(t_0)=\wh{b}.
\ee
At time $t_1$\/, after the measurement has taken place,
\be
\wh{g}(t_1)=\wh{U}\da\wh{g}\wh{U}, \label{UgU}
\ee
\be
\wh{b}(t_1)=\wh{U}\da\wh{b}\wh{U}. \label{UbU}
\ee
Using (\ref{U})-(\ref{P_S}), (\ref{UgU}) and (\ref{UbU}), we find that
\be
\wh{g}(t_1)=\wh{g},
\ee
\be
\wh{b}(t_1)=\sum_{i=1}^L \wh{b}_i \otimes \wh{P}_i^{\wti{S}} \label{b_t_1}
\ee
where
\be
\wh{b}_i=\wh{u}_i\da \wh{b} \wh{u}_i,
\ee
so
\be
\wh{b}_i|\cO;\bet_0\ra=\bet_i|\cO;\bet_0\ra.
\ee
The fact that at time $t_1$\/ the operator  $\wh{b}(t)$\/ takes the form (\ref{b_t_1}),
a sum of operators in the state space of $\cO$\/ each 
of which is ``labeled''
with a factor acting in the state space of $\cS$\/,
is taken to indicate that the state of awareness of $\cO$\/ at that time
can  
be regarded,  for all practical purposes, as split into  $L$\/ noninteracting copies,
with copy $i$\/ perceiving measurement result $\bet_i$\/\cite{Rubin01}. We will term this
``interpretive rule 1.''

To this interpretive rule we adjoin the following ``interpretive rule 2'':
At any time $t$\/ only those copies of $\cO$\/ exist which
have nonzero values for $W_i(t)$\/,  
\be
W_i(t) \neq 0,
\ee
where the ``weight'' $W_i(t)$\/ is defined
as the matrix element of the label factor between the  initial 
state (\ref{uncorrelated_state}) and its adjoint; in the present example,
\be
W_i(t) \equiv \la  \psi (t_0) | \wh{P}_i^{\wti{S}} | \psi (t_0) \ra
= \la \cS; \psi  | \wh{P}_i^{\wti{S}} |\cS; \psi  \ra
. \label{weight_def}
\ee
The condition
\be
W_i(t)=0,  \label{zero_weight}
\ee
indicating that no observer-copy exists at time $t$\/ who perceives measurement result
$\bet_i$\/
will play a role analogous to the condition (\ref{zero_norm}) in 
the approaches of  DeWitt and Okhuwa---with the distinction that condition  (\ref{zero_weight}),
for each value of $i$\/,
explicitly refers to a {\em single} Everett copy of an observer.

\section{The Frequency Interpretation of Probability and the Born Rule in Everett Quantum Mechanics} \label{sec_FI}

\subsection{State Spaces of Systems and Observers}

We consider an ensemble of $N$\/ identical physical systems $\cS^{(p)}$\/, $p=1,\ldots, N$.\/
The state space of $\cS^{(p)}$\/ is  spanned by the two eigenstates of an operator
$\wh{a}^{(p)}$\/ which acts nontrivially only in that state space:
\be
\wh{a}^{(p)} |\cS^{(p)}; \al_{i^{(p)}} \ra =\al_{i^{(p)}}|\cS^{(p)};\al_{i^{(p)}} \ra,
\hs i^{(p)}=1,2, \hs  p=1,\ldots, N,
\ee
\be 
\al_{1} \neq  \al_{2}.
\ee
To each system $\cS^{(p)}$\/ there corresponds an observer/measuring device $\cO^{(p)}$\/,
the state space of which is spanned by the three eigenvectors of an operator
$\wh{b}\pup$\/ which acts nontrivially only in that state space:
\be
\wh{b}^{(p)} |\cO^{(p)}; \bet_{i^{(p)}} \ra =\bet_{i^{(p)}}|\cO^{(p)};\bet_{i^{(p)}} \ra,
\hs i^{(p)}=0,1,2, \hs  p=1,\ldots, N,
\ee
\be 
\bet_{i} \neq  \bet_{j}, \hs i \neq j.
\ee
The observer $\cO\pup$\/ interacts with the system $\cS\pup$\/ in such a manner as to
determine the value of the observable $\wh{a}\pup$\/, as described in the previous section.
The eigenvalue $\bet_0$\/ indicates the state of ignorance, while $\bet_1$\/ and
$\bet_2$\/ correspond, respectively, to  $\cO\pup$\/ having measured $\al_1$\/ or
$\al_2$\/.  

After these $N$\/ measurement interactions have taken place---it is immaterial whether
they take place simultaneously or in any arbitrary order since, by virtue of the
fact that each measurement interaction affects a distinct system-observer 
pair, the operators corresponding to the different interactions 
commute---an additional observer ${\cal F}$\/ interacts with all of the $\cO\pup$\/ 
so as to determine the relative frequency of the result $\bet_1$\/ among the  observers.
Since there are $N$\/ observers, the possible values of a measurement
of the relative frequency are the $N+1$\/ numbers
0, $1/N$\/, $2/N$\/, \ldots, 1.    As $N \rightarrow \infty$\/ the
number of possible values for the relative frequency grows without bound,
and the difference between adjacent values shrinks to zero.

As emphasized in this context by Graham\cite{Graham73},
the resolution of any real measuring
device is finite.
Suppose $\cF$\/  queries each observer in sequence and in the end computes the relative frequency. If the physical device $\cF$\/ occupies a
finite volume and has at its disposal a finite amount of
energy, it will be able to record at most a finite number of measurements.
Suppose, for example, it uses the directions of spins of neutrons to record
the responses of each observer $\cO\pup$\/. As more neutrons are
added to the finite volume available to $\cF$\/, more energy is required
to pack the neutrons into the volume as a result of
degeneracy pressure.  An  upper limit on how
much information can be recorded in this manner is set by the number
of neutrons which, when located within the finite available volume,
would have sufficient mass to produce a black hole with an event horizon
surrounding the volume. (The black-hole
limit would, of course, apply as well to any information-storage
scheme involving bosonic degrees of freedom. See, e.g., \cite{Beckenstein01}
and references therein.) Of course, existing information-storage
devices have capacities well below these limits\cite{Beckenstein01}. As 
$N \rightarrow \infty$\/, 
 $\cF$\/
will have to either stop recording new information from the $\cO\pup$\/'s
or discard old information. The situation is qualitatively
no different if $\cF$\/  updates
the value of the relative frequency after interacting with each observer.
As $N \rightarrow \infty$\/ the number of digits required to record
any possible number of the form (integer from 0 through $N$\/)$\times 1/N$\/ will grow until
it outstrips the available memory.

So, we take the state space of $\cF$\/ to be spanned by  $\nu +2$\/ eigenvectors:
\be
\wh{f}|\cF;\phi_i\ra =\phi_i |\cF;\phi_i\ra, \hs i=0,\ldots,\nu+1,
\ee
where, for $i > 0$\/, the eigenvalue $\phi_i$\/ is one of the possible
outputs of $\cF$\/ after it has interacted with all of the $\cO\pup$\/'s,
\be
\phi_i= (i-1)/\nu, \hs  i=1,\ldots, \nu+1. \label{phi_i_val}
\ee
$\phi_0$\/ will be taken to indicate a state of ignorance. For this purpose 
the only requirement is that $\phi_0$\/ not equal any of the other eigenvalues
(\ref{phi_i_val}); it will be 
convenient to assign it the value
\be
\phi_0=-1/\nu.\label{phi_0_val}
\ee

\subsection{Measurement Interactions}

\subsubsection{Measurement of $\cS\pup$\/ by $\cO\pup$\/}

The unitary operator corresponding to the interaction between
$\cS\pup$\/ and $\cO\pup$\/, following Sec. \ref{meas_schrod_pic}
above, is
\be
\wh{U}\pup=\sum_{i=1}^2 \wh{u}_i\pup \otimes \wh{P}_i^{\cS\pup}, \hs p=1, \ldots, N,
\label{U_p_def}
\ee
where
\be
\wh{P}_i^{\cS\pup}=|\cS\pup; \al_i \ra\la\cS\pup; \al_i |, \hs i=1,2, \hs p=1, \ldots, N,
\label{P_cS_p_def}
\ee
and
\be
\wh{u}_i\pup |\cO\pup;\bet_0\ra= |\cO\pup;\bet_i\ra, \hs i=1,2, \hs p=1, \ldots, N.
\label{u_p_action}
\ee
From (\ref{U_p_def}), (\ref{P_cS_p_def}) and the unitarity of $\wh{U}\pup$\/ it follows that
\be
\wh{u}_i^{(p)\dag}\wh{u}_i^{(p)}=1, \hs i=1,2, \hs p=1, \ldots, N. \label{u_i_p_orthogonal}
\ee

\subsubsection{Measurement of $\cO\pup$\/   by  $\cF$\/}

Define the relative frequency function for the result $\bet_{i\pup}=\bet_{1}$\/,
\be
f(\bet_{i^{(1)}}, \ldots, \bet_{i^{(N)}})=(1/N)\sum_{p=1}^N \delta_{i\pup,1},
\hs i\pup \neq 0 \hs \forall \hs p.
\label{rel_freq_func}
\ee
The possible values of this function are
\be
f_l=l/N, \hs l=0,\ldots,N. \label{f_l_vals}
\ee
Define also the finite-resolution relative frequency function
$\phi(\bet_{i^{(1)}}, \ldots, \bet_{i^{(N)}})$\/
to be that $\phi_i$\/, $i=1, \ldots, \nu+1$\/, which is closest in
value to  $f(\bet_{i^{(1)}}, \ldots, \bet_{i^{(N)}})$\/:
\begin{eqnarray}
\phi(\bet_{i^{(1)}}, \ldots, \bet_{i^{(N)}})&=&
\arg\min_{\phi_i} |\phi_i -f(\bet_{i^{(1)}}, \ldots, \bet_{i^{(N)}})|, \label{fin_res_freq_fun}
\hs i\pup \neq 0 \hs \forall \hs p\\
&& \hbox{\rm (smaller $\phi_i$\/ in case of a tie)} \label{case_of_tie} \nonumber
\end{eqnarray}
where the $\phi_i$\/'s are as given in (\ref{phi_i_val}).
It will also prove convenient to define
\begin{eqnarray}
\wti{\phi}(\bet_{i^{(1)}}, \ldots, \bet_{i^{(N)}}) &=& \phi(\bet_{i^{(1)}}, \ldots, \bet_{i^{(N)}}), \hs i\pup \neq 0 \hs \forall \hs p\\
&=& \phi_0 \hs \mbox{\rm otherwise} \label{phi_0_otherwise}
\end{eqnarray}
where $\phi_0$\/ is as given in (\ref{phi_0_val}).

The unitary operator corresponding to the measurement of all of the $\cO\pup$\/'s
by $\cF$\/ can then be written as
\be
\wh{U}_{\cF}=\sum_{k=0}^{\nu+1}\wh{u}_k^{\cF}\otimes \wh{P}_k^{\wti{\cO}},\label{U_F}
\ee
where
\be
\wh{u}_k^{\cF}|\cF;\phi_0\ra=|\cF;\phi_k\ra, \hs k=0, \ldots, \nu + 1,
\ee
Note that $\wh{u}_0^{\cF}$\/ acts as the identity on the ignorant state
$|\cF;\phi_0\ra$\/. $\wh{P}_k^{\wti{\cO}}$\/ is the
projection operator which projects out those states corresponding
to $\phi_k$\/:
\be 
\wh{P}_k^{\wti{\cO}}=\sum_{i^{(1)}=0}^2 \ldots \sum_{i^{(N)}=0}^2 
\delta_{\nu \wti{\phi}(\bet_{i^{(1)}}, \ldots, \bet_{i^{(N)}}), k-1}
\otimes_{p=1}^N \wh{P}_{i^{(p)}}^{\cO^{(p)}} \label{P_wticO_def}
\ee
where
\be
\wh{P}_{i^{(p)}}^{\cO^{(p)}}=|\cO\pup;\bet_{i\pup}\ra \la \cO\pup;\bet_{i\pup}|.
\label{P_cO_i_proj}
\ee

Using (\ref{phi_i_val}), (\ref{phi_0_val}) and (\ref{fin_res_freq_fun})-(\ref{phi_0_otherwise})
we see that, for any given set of values for $i^{(1)}, \ldots, i^{(N)}$\/,
the quantity $\nu \wti{\phi}(\bet_{i^{(1)}}, \ldots, \bet_{i^{(N)}})$\/ is
equal to one of the values $-1, 0, 1, \ldots, \nu$\/. Therefore
\be
\delta_{\nu \wti{\phi}(\bet_{i^{(1)}}, \ldots, \bet_{i^{(N)}}), k-1}
\delta_{\nu \wti{\phi}(\bet_{i^{(1)}}, \ldots, \bet_{i^{(N)}}), l-1}
=\delta_{k,l}\delta_{\nu \wti{\phi}(\bet_{i^{(1)}}, \ldots, \bet_{i^{(N)}}), k-1}
\label{delta_prod}
\ee
and
\be
\sum_{k=0}^{\nu+1}\delta_{\nu \wti{\phi}(\bet_{i^{(1)}}, \ldots, \bet_{i^{(N)}}), k-1}
=1. \label{delta_sum}
\ee
Using these we verify that
\be
\wh{P}_k^{\wti{\cO}}\wh{P}_l^{\wti{\cO}}=\delta_{k,l}\wh{P}_k^{\wti{\cO}}, \hs
k,l=0,\ldots,\nu + 1,
\label{P_wticO_orthogonal}
\ee
and
\be
\sum_{k=0}^{\nu + 1}\wh{P}_k^{\wti{\cO}}= 1.
\label{P_wticO_complete}
\ee
From (\ref{U_F}), (\ref{P_wticO_orthogonal}) and the unitarity of $\wh{U}_{\cF}$\/ it follows that
\be
\wh{u}_k^{\cF \dag}\wh{u}_k^{\cF}=1, \hs k=0,\ldots,\nu+1.
\ee

\subsubsection{Complete Measurement Transformation Operator}

The unitary operator corresponding to all the $\cO\pup$\/'s 
measuring their $\cS\pup$\/'s, followed by $\cF$\/ measuring all the $\cO\pup$\/'s and determining the relative frequency
of $\bet_1$\/ observations, is
\be
\wh{U}=\wh{U}_{\cF} \wh{U}_{\cO}, \label{complete_U}
\ee
where
\be
\wh{U}_{\cO}=\otimes_{p=1}^N \wh{U}\pup. \label{U_cO}
\ee

\subsection{Post-Measurement Heisenberg-Picture Operators}

\subsubsection{{\boldmath $ \wh{a}\pup(t_1)$\/}} \label{subsubsec_a}

After measurement,
\be
\wh{a}\pup(t_1)=\wh{U}\da \; \wh{a}\pup\ \; \wh{U}, \label{a_t1}
\ee
or, using (\ref{complete_U}),
\be
\wh{a}\pup(t_1)=\wh{U}_\cO\da \; \wh{U}_{\cF}\da \; \wh{a} \; \wh{U}_{\cF}\; \wh{U}_{\cO}.
\label{a_t1_2}
\ee
From (\ref{U_F})-(\ref{P_cO_i_proj}) we see that $\wh{U}_{\cF}$\/ doesn't act in the state space
of $\cS\pup$\/, so
\be
[\wh{U}_{\cF},\wh{a}\pup]=0,
\ee
and (\ref{a_t1_2}) can be written as
\begin{eqnarray}
\wh{a}\pup(t_1)&=&\wh{U}_{\cO}\da \; \wh{a}\pup \; \wh{U}_{\cO}\\
               &=&\left(\otimes_{q=1}^N \wh{U}^{{(q)}\dag} \right)\; 
                  \wh{a} \; 
                  \left(\otimes_{r=1}^N \wh{U}^{(r)} \right) \label{a_t1_3}
\end{eqnarray}
using (\ref{U_cO}). From (\ref{U_p_def})and (\ref{P_cS_p_def}) we see that  $\wh{U}^{(q)}$\/ only acts nontrivially 
on $\wh{a}\pup$\/ for $q=p$\/, so,  using  (\ref{U_p_def}), (\ref{a_t1_3}) becomes
\be
\wh{a}\pup(t_1)=
\left(\sum_{i^{(p)}=1}^2 \wh{u}_{i^{(p)}}^{(p)\dag} \otimes \wh{P}_{i^{(p)}}^{\cS^{(p)}}\right)\;
\wh{a}\pup \;
\left(\sum_{j^{(p)}=1}^2 \wh{u}_{j^{(p)}}^{(p)} \otimes \wh{P}_{j^{(p)}}^{\cS^{(p)}}\right).
\label{a_t1_4}
\ee
Since
\be
\wh{a}\pup = \sum_{k\pup=1}^2 \al_{k\pup} \wh{P}_{k\pup}^{\cS\pup},
\ee
it follows from (\ref{a_t1_4}), (\ref{P_cS_p_def}) and (\ref{u_i_p_orthogonal}) that
$\wh{a}\pup(t)$\/ is unchanged by the measurement process,
\be
\wh{a}\pup(t_1)=\wh{a}\pup.
\ee

\subsubsection{{\boldmath $\wh{b}\pup(t_1)$\/}}\label{subsubsec_b}

After measurement,
\be
\wh{b}\pup(t_1)=\wh{U}\da \; \wh{b}\pup\ \; \wh{U}. \label{b_t1}
\ee
Using (\ref{complete_U}),
\be
\wh{b}\pup(t_1)=\wh{U}_\cO\da \; \wh{U}_{\cF}\da \; \wh{b}\pup \; \wh{U}_{\cF}\; \wh{U}_{\cO},
\label{b_t1_2}
\ee
or
\be
\wh{b}\pup(t_1)=\wh{U}_\cO\da \; \wh{X}_b\pup \wh{U}_{\cO}, \label{UXbU}
\ee
where the intermediate quantity $\wh{X}_b\pup$\/ is defined as
\be
\wh{X}_b\pup=\wh{U}_{\cF}\da \; \wh{b}\pup \; \wh{U}_{\cF}. \label{X_b_def}
\ee
From (\ref{X_b_def}) and (\ref{U_F}),
\begin{eqnarray}
\wh{X}_b\pup&=&\left(\sum_{k=0}^{\nu+1}\wh{u}_k^{\cF}\otimes \wh{P}_k^{\wti{\cO}}\right)
         \wh{b}\pup
         \left(\sum_{l=0}^{\nu+1}\wh{u}_l^{\cF}\otimes \wh{P}_l^{\wti{\cO}}\right)\\
&=&\sum_{k,l=0}^{\nu + 1}\sum_{i=0}^{2}\wh{u}_k^{\cF\dag}\wh{u}_l^{\cF}\otimes \bet_i
\wh{P}_k^{\wti{\cO}}\wh{P}_i^{\cO\pup}\wh{P}_l^{\wti{\cO}} \label{X_b_3}
\end{eqnarray}
since
\be
\wh{b}\pup=\sum_{i=0}^2 \bet_i \wh{P}_i^{\cO\pup}. \label{spectral_b}
\ee
Using (\ref{P_wticO_def})-(\ref{delta_prod})
\be
\wh{P}_k^{\wti{\cO}}\wh{P}_i^{\cO\pup}\wh{P}_l^{\wti{\cO}}=
\delta_{k,l} \sum_{i^{(1)}=0}^2 \ldots \sum_{i^{(N)}=0}^2 \delta_{i^{(p)},i}\;
\delta_{\nu \wti{\phi}(\bet_{i^{(1)}}, \ldots, \bet_{i^{(N)}}), k-1} \otimes_{q=1}^N \wh{P}_{i^{(q)}}^{\cO^{(q)}}. \label{tri_P_prod}
\ee
Using (\ref{tri_P_prod}), (\ref{delta_prod}) and (\ref{spectral_b}) in (\ref{X_b_3}),
\be
\wh{X}_b\pup=\wh{b}\pup. \label{X_b_val}
\ee
Using (\ref{X_b_val}), (\ref{U_p_def}),  (\ref{P_cS_p_def}),  (\ref{u_i_p_orthogonal}) and 
(\ref{U_cO})  in (\ref{UXbU}),
\be
\wh{b}^{(p)}(t_1)=\sum_{i=1}^2 \wh{b}_i^{(p)}\otimes \wh{P}_i^{\cS\pup}, \label{b_t2_val}
\ee
where
\be
\wh{b}_i^{(p)}=\wh{u}_i^{(p)\dag} \wh{b}^{(p)} \wh{u}_i^{(p)}.
\ee

\subsubsection{{\boldmath $\wh{f}(t_1)$\/}}\label{subsubsec_f}

After measurement,
\be
\wh{f}(t_1)=\wh{U}\da \; \wh{f} \; \wh{U}. \label{f_t1}
\ee
Using (\ref{complete_U}),
\be
\wh{f}(t_1)=\wh{U}_\cO\da \; \wh{U}_{\cF}\da \; \wh{f} \; \wh{U}_{\cF}\; \wh{U}_{\cO},
\label{f_t1_2}
\ee
or
\be
\wh{f}(t_1)=\wh{U}_\cO\da \; \wh{X}_f \wh{U}_{\cO}, \label{UXfU}
\ee
where the intermediate quantity $\wh{X}_f$\/ is defined as
\be
\wh{X}_f=\wh{U}_{\cF}\da \; \wh{f} \; \wh{U}_{\cF}. \label{X_f_def}
\ee
Using (\ref{U_F}) and (\ref{P_wticO_orthogonal}), (\ref{X_f_def}) becomes
\be
\wh{X}_f=\sum_{k=0}^{\nu+1} \wh{u}_k^{\cF\dag} \; \wh{f} \; \wh{u}_k^{\cF}
\otimes \wh{P}_k^{\wti{\cO}}.  \label{X_f_val}
\ee
Using this in (\ref{UXfU}), 
\be
\wh{f}(t_1)=\sum_{k=0}^{\nu + 1} \wh{f}_k \otimes \wh{L}_k,\label{f_k_t1}
\ee
where
\be
\wh{f}_k=\wh{u}_k^{\cF\dag} \;  \wh{f} \; \wh{u}_k^{\cF}, \hs k=0,\ldots,\nu+1,
\ee
and
\be
\wh{L}_k=\sum_{i^{(1)}=1}^2 \ldots \sum_{i^{(N)}=1}^2 
\left(
\left(\otimes_{p=1}^N \wh{u}_{i^{(p)}}^{(p)\dag}\right) \wh{P}_k^{\wti{\cO}} 
\left(\otimes_{q=1}^N \wh{u}_{i^{(q)}}^{(q)}\right)
\right)
\otimes_{r=1}^N \wh{P}_{i^{(r)}}^{\cS^{(r)}}, \hs k=0,\ldots,\nu+1. \label{L_k_def}
\ee

\subsection{Initial State}

For the constant Heisenberg-picture state we take the product
state in which $\cF$\/ and the $\cO\pup$\/'s are ignorant and
each of the $\cS\pup$\/'s is in a superposition of $|\cS\pup;\al_1\ra$\/
and $|\cS\pup;\al_2\ra$\/ with the same coefficients:
\be
|\psi(t_0)\ra= |\cF;\phi_0 \ra \prod_{p=1}^N |\cO\pup;\bet_0\ra
				       \prod_{q=1}^N |\cS^{(q)};\psi_0\ra, \label{big_init_state}
\ee
where
\be
|\cS^{(q)};\psi_0\ra = c_1^{(q)}|\cS^{(q)};\al_1\ra + c_2^{(q)}|\cS^{(q)};\al_2\ra,
\hs q=1,\ldots,N, \label{Sp_init_state}
\ee
with
\be
c_1^{(q)}=c_1,\hs c_2^{(q)}=c_2 \hs \forall \; q, \label{c_same}
\ee
\be
|c_1|^2 + |c_2|^2=1. \label{c_sum}
\ee

\subsection{Weights} \label{subsec_W}

Using (\ref{big_init_state}), the definition of the weight
 in Sec. \ref{subsec_HP}  and the results of Secs. \ref{subsubsec_a} - \ref{subsubsec_f}
we see that the weights associated with $\wh{a}\pup$\/ both before and after
measurement, and with $\wh{b}\pup$\/ and $\wh{f}$\/ before measurement, are simply
unity.  All of these operators are of the form (\ref{b_t_1}) only in a trivial sense.

After measurement, $\wh{b}\pup$\/ and $\wh{f}$\/ are nontrivially in the form (\ref{b_t_1}).
The weight associated with $\wh{b}_i\pup$\/  is, 
according to (\ref{b_t2_val}),
\be
W_{b,i}\pup(t_1)=\la\psi(t_0)|\wh{P}_i^{\cS\pup}|\psi(t_0)\ra, \hs i=1,2, \hs p=1,\ldots,N.
\ee
Using (\ref{P_cS_p_def}) and 
(\ref{big_init_state})-(\ref{c_same}), we find
\be
W_{b,i}\pup(t_1)=|c_i|^2, \hs i=1,2, \hs p=1,\ldots,N,
\ee
which is nonzero for any value of $N$\/ and for both values of $i$\/,
unless either $c_1=0$\/ or $c_2=0$\/.

From (\ref{f_k_t1}), the weight for $\wh{f}_k$\/ is
\be
W_{f,k}(t_1)=\la \psi(t_0)| \wh{L}_k | \psi(t_0)\ra.
\ee
Using (\ref{L_k_def}), (\ref{big_init_state})-(\ref{c_same}), (\ref{u_p_action}),
(\ref{P_wticO_def}), and (\ref{P_cO_i_proj}),
\be
W_{f,k}(t_1)=\sum_{i^{(1)}=1}^2 \ldots \sum_{i^{(N)}=1}^2\delta_{\nu\wti{\phi}(\bet_{i^{(1)}}, \ldots, \bet_{i^{(N)}}),k-1}
\prod_{r=1}^N |c_{i^{(r)}}|^2, \hs k=0,\ldots,\nu+1.
\ee
From (\ref{phi_i_val}), (\ref{phi_0_val}), and 
(\ref{fin_res_freq_fun})-(\ref{phi_0_otherwise}), we see that, for
any value of $N$\/, $W_{f,k}(t_1)$\/ vanishes for $k=0$\/,
\be
W_{f,0}(t_1)=0 \hs \forall \; N.
\ee
So, at $t_1$\/, there is no Everett copy of the relative-frequency observer $\cF$\/
who has the perception of being in a state of ignorance.  This is of course what
we expect to find, given that we have defined the evolution operators $\wh{U}\pup$\/
to measure the states of the $\cS\pup$\/'s without error, and the evolution operator
$\wh{U}_{\cF}$\/ to accurately determine the $\cO\pup$\/'s  measurements and
compute the appropriate finite-resolution frequency $\phi_i$\/.

For $k=1,\ldots,\nu+1$\/,
\be
W_{f,k}(t_1)=\sum_{i^{(1)}=1}^2 \ldots \sum_{i^{(N)}=1}^2\delta_{\nu \phi(\bet_{i^{(1)}}, \ldots, \bet_{i^{(N)}}),k-1}
\prod_{r=1}^N |c_{i^{(r)}}|^2. \label{W_f_k_intermediate}
\ee
Using (\ref{fin_res_freq_fun})-(\ref{phi_0_otherwise}), (\ref{delta_sum}), (\ref{c_same}),
(\ref{c_sum}) and the binomial theorem,  we
verify that
\be
\sum_{k=1}^{\nu+1}W_{f,k}(t_1)=1. \label{W_f_k_sum_1}
\ee

Using (\ref{phi_i_val}) and (\ref{rel_freq_func})-(\ref{phi_0_otherwise}) 
in (\ref{W_f_k_intermediate}), we obtain
\be
W_{f,k}(t_1)=\sum_{l \; \; | \; \; 0 \leq l \leq N, \;  N(\phi_k-(1/2\nu)) < l \leq N(\phi_k + (1/2\nu))}
p_{N,l}\label{W_f_k}
\ee
where
\be
p_{N,l}=\frac{N!}{l!(N-l)!}|c_1|^{2l}|c_2|^{2(N-l)}. \label{p_N_l_def}
\ee
To evaluate (\ref{W_f_k}) in the limit $N \rightarrow \infty$\/ we
make use of Bernoulli's law of large numbers. From \cite[p.195]{Renyi70}, for example,
we have, in our notation,
\be
\lim_{N \rightarrow \infty} S_N=0,\label{weak_law_0}
\ee
where
\be
S_N= \sum_{l \;\; | \;\; 0 \leq l \leq N, \; |l-N|c_1|^2| > N\eps} p_{N,l}. 
\label{weak_law}
\ee
(The value of $\eps$\/
in (\ref{weak_law}) is  independent of $N$\/.) 
This is true for every positive number $\eps$\/.  From ({\ref{p_N_l_def})
we see that
\be
p_{N,l} \ge 0 \; \forall \; N,l.
\ee
Therefore, in 
(\ref{weak_law}),  we can, for each $N$\/, 
replace the sum  
over all integers $l$\/ between $0$\/ and $N$\/ inclusive satisfying $|l-N|c_1|^2| > N\eps$\/ with the sum over an arbitrary subset $\{l\}_N$\/  of these integers. That is, 
\be
\lim_{N \rightarrow \infty} \wti{S}_N=0,\label{weaker_law_0}
\ee
where
\be 
\wti{S}_N= \sum_{\{l\}_N \;\; | \;\;0 \leq l \leq N, \; |l-N|c_1|^2| > N\eps} p_{N,l},
\label{weaker_law}
\ee
since $|\wti{S}_N| \leq |S_N| \; \forall \; N$\/.

Suppose $|c_1|^2$\/ is closer
to the finite resolution relative frequency $\phi_{k\pr}$\/ than to any other
$\phi_k$\/, $k=1,\ldots,\nu + 1$\/,
\be
k\pr=\arg \min_k |\phi_k-|c_1|^2|.
\ee
For concreteness say that $|c_1|^2$\/  is less
than $\phi_{k\pr}$\/,
\be
|c_1|^2=\phi_{k\pr}-\Delta, \label{c_1_Delta}
\ee
where
\be
0 < \Delta < \frac{1}{2\nu}. \label{Delta_bounds}
\ee
(See Fig. \ref{Fig1}.)
\begin{figure}
\centering
\underline{$| \leftarrow \frac{1}{2\nu} \rightarrow |\leftarrow \frac{1}{2\nu} \rightarrow |
\leftarrow \frac{1}{2\nu} \rightarrow |\leftarrow \frac{1}{2\nu} \rightarrow |$}\\
\vspace*{1mm}
$\hspace*{.74in}\phi_{k\ppr} \hspace*{.66in}|c_1|^2 \hspace*{.25in}\phi_{k\pr} \hspace*{.65in}$\\
$\hspace*{1.405in} | \hspace{-.15cm} \leftarrow \hspace{-.1cm} \Delta \hspace{-.1cm} \rightarrow \hspace{-.15cm}|
\hspace*{.4825in}$\\
\caption{Finite resolution relative frequencies $\phi_{k\ppr}$\/, 
$\phi_{k\pr}$\/ and Born-rule  probability $|c_1|^2 $\/ 
in the case that $|c_1|^2  < \phi_{k\pr}$\/.}
\label{Fig1}
\end{figure}
Consider now the expression (\ref{W_f_k}) for the weight $W_{f,k\ppr}(t_1)$\/, where
\be
k\ppr=k\pr-1. \label{kppr}
\ee
The values of $l$\/ which enter  into the sum in (\ref{W_f_k}), for $k=k\ppr$\/,  are bounded above
by
\be
l_{ub}=N\left(\phi_{k\ppr}+\frac{1}{2\nu}\right). \label{ub}
\ee
Therefore, using (\ref{phi_i_val}), (\ref{c_1_Delta}), (\ref{kppr}) and  (\ref{ub}),
 the values that the quantity $|l-N|c_1|^2|$\/ can
take for any $l$\/ appearing in the sum in $W_{f,k\ppr}(t_1)$\/ are bounded below by
\be
|l_{ub}-N|c_1|^2|=N\left(\frac{1}{2\nu}-\Delta\right).
\ee 
So, for any positive $\eps$\/ such that
\be
\eps  < \left(\frac{1}{2\nu}-\Delta\right)
\ee
the sum in (\ref{W_f_k}) is of the form  (\ref{weaker_law}), 
implying, by (\ref{weaker_law_0}),
\be
\lim_{N \rightarrow \infty} W_{f,k\ppr}(t_1)=0.
\ee
For   $k \neq k\ppr$\/ or $k\pr$\/, allowed values of $l$\/ in the sum  
in the expression (\ref{W_f_k}) for $W_{f,k}(t_1)$\/ are bounded away from $N|c_1|^2$\/ even more strongly, so the same argument applies, as it  does  also
in the case that $|c_1|^2 > \phi_{k\pr}$\/. So,  we conclude that
\be
\lim_{N \rightarrow \infty} W_{f,k}(t_1)=0, \hs k \neq k\pr. \label{W_f_k_0}
\ee

The above analysis does not apply to $W_{f,k\pr}(t_1)$\/ since, for fixed $\eps$\/,
the sum will include values of $l$\/ approaching closer to $N|c_1|^2$\/ than
$N\eps$\/ for $N$\/ sufficiently large, no matter how small a fixed
value of $\eps$\/ is  used. From (\ref{W_f_k_sum_1}) and  (\ref{W_f_k_0})
we see that
\be
\lim_{N \rightarrow \infty} W_{f,k\pr}(t_1)=1.
\ee

In the case that $|c_1|^2$\/ is precisely equidistant from two adjacent $\phi_k$\/'s,
\be
|c_1|^2=\phi_{k^<} + \frac{1}{2\nu} = \phi_{k^>} - \frac{1}{2\nu},
\ee
\be
k^>=k^<+1,
\ee
the argument  above goes through for all $k\neq k^< \;  \mbox{\rm or} \; k^>$\/ to show that
\be
\lim_{N \rightarrow \infty} W_{f,k}(t_1)=0, \hs k\neq k^< \;  \mbox{\rm or} \; k^>.
\ee
Using (\ref{W_f_k_sum_1}), we therefore conclude
\be
\lim_{N \rightarrow \infty} \left(W_{f,k^<}(t_1) +  W_{f,k^>}(t_1) \right) =1.
\label{sum_Ws_even}
\ee
In the case that $|c_1|^2 = |c_2|^2 =1/2$\/ and $\nu$\/ is odd, 
the sums (\ref{W_f_k}) for 
$W_{f,k^<}(t_1)$\/
and for $W_{f,k^>}(t_1)$\/ are term-by-term identical  except for  one term:
\be
W_{f,k^<}(t_1)=T(N)+T^<(N), \label{W_left_sum}
\ee
\be
W_{f,k^>}(t_1)=T(N)+T^>(N),\label{W_right_sum}
\ee
where
\be
T(N)= \sum_{l \; \;| \; \;  N\left(\frac{\nu-2}{2\nu}\right) < l < \frac{N}{2}}
\frac{N!}{2^{N}l! (N-l)!},
\ee
\begin{eqnarray}
T^<(N)&=&\frac{N!}{2^{N}\left(\left[ \frac{N}{2}\right] ! \right)^2}, \hs N \; \;  
\mbox{\rm even,}\\
&=& 0  \hspace*{1in} \mbox{\rm otherwise,}
\label{term1}
\end{eqnarray}
and
\begin{eqnarray}
T^>(N)&=& \frac{N! }{2^{N}\left[N\left(\frac{\nu-2}{2\nu}\right)\right]!
                 \left[N\left(\frac{\nu+2}{2\nu}\right)\right]!}, \; \; 
N\left(\frac{\nu+2}{2\nu}\right) \; \; \mbox{\rm is an integer $>$\/ 0}\\
&=& 0 \hspace*{2in} \mbox{\rm otherwise.}
\label{term2}
\end{eqnarray}
From  Stirling's
formula, 
\be
\lim_{N \rightarrow \infty} T^<(N)=\lim_{N \rightarrow \infty} T^>(N)= 0. \label{T_T_lim}
\ee
Using  (\ref{sum_Ws_even})-(\ref{W_right_sum}) and 
(\ref{T_T_lim}), 
\be
\lim_{N \rightarrow \infty} T(N)=1/2. \label{T_lim}
\ee
(The limit of a sum of sequences is the sum of the limits of those
sequences; see, e.g., 
\cite[p. 49]{Rudin76}).  We conclude from (\ref{W_left_sum}), (\ref{W_right_sum}),
(\ref{T_T_lim}) and (\ref{T_lim}) that
\be
\lim_{N \rightarrow \infty} W_{f,k^<}(t_1) = \lim_{N \rightarrow \infty} W_{f,k^>}(t_1) 
=1/2 \neq 0.
\ee

So, in the limiting case  of an infinitely large ensemble of identical systems
measured by an observer $\cF$\/ performing a real (finite-resolution) relative-frequency measurement
there is, after measurement, either a single Everett copy of $\cF$\/ who perceives that finite-resolution 
relative frequency $\phi_k$\/ closest to the Born-rule value $|c_1|^2$\/,
or two Everett copies of $\cF$\/ 
who respectively perceive one of the two $\phi_k$\/'s
closest to $|c_1|^2$\/.

\section{Summary}\label{sec_summary}

\noindent
Premises:
\begin{enumerate}
\item Definition of measurement situation:
	\begin{enumerate}
	\item The initial state of observer and observed system is of the uncorrelated form
             (\ref{uncorrelated_state}).
	\item The interaction between observer and system leaves the Heisenberg-picture operator corresponding to the state of belief of the observer
            in the form (\ref{b_t_1}).
	\item The observer has finite resolution, i.e., is capable of perceiving a discrete finite set of possible outcomes.
	\end{enumerate}
\item Interpretation of operators after measurement:
	\begin{enumerate}
	\item Each of the terms in (\ref{b_t_1}) corresponds to a copy of the observer with
            a distinct perception of the outcome of the measurement; with the proviso that
      \item Only those copies with nonzero weight (see Sec. \ref{subsec_HP}) exist.
	\end{enumerate}
\end{enumerate}
Conclusions:
\begin{enumerate}
\item After the measurement of relative frequency, in the limiting case of an infinitely large ensemble of measured systems, the only existing Everett
copies of an observer measuring
relative frequency will perceive results equal to the Born rule values, to within
available resolution.
\item If one defines probability in accord with the frequency interpretation (see Sec. \ref{sec_intro}), then probability exists in Everett quantum mechanics, in that both
randomness and, given any particular experiment (as specified by the initial state), an essentially  unique limiting value for the measured relative frequency of an outcome in an ensemble
of identical experiments, exist.
\end{enumerate}

\section{Discussion} \label{sec_discussion}

The above analysis has been limited to ``measurement situations.''\footnote{These situations
are, of course, taken to be simply those which happen to satisfy certain physical criteria. They
remain subject to the unitary quantum-mechanical evolution laws, as distinct from
the case in the Copenhagen interpretation.} 
It may be possible to generalize it 
to apply to a broader class of situations.
However, it is not necessary that it be possible to generalize it to  apply to all possible physical situations.
It is a hallmark of quantum mechanics that in many situations, for many
observables, probability
 simply is not defined\cite{Griffiths84,Hartle95}. 
(Page 
argues that probabilities need only be defined in the context of situations involving
conscious perceptions\cite{Page94,Page95,Page96}.)

The ontology of the Everett interpretation in the Heisenberg picture is 
different from that in the Schr\"{o}dinger picture. Consider the scenario of
Sec. \ref{sec_FI} above, and,  for the moment, the case of finite $N$\/.
The Everett interpretation in the Schr\"{o}dinger picture
would describe the situation after all the measurements have been
made in terms of $2^N$\/ branches of the state vector.
In the Heisenberg picture, there are two copies of each of the $N$\/ systems $\cS\pup$\/,
two copies of each of the $N$\/ observers $\cO\pup$\/, and $\nu+1$\/
copies of $\cF$\/. The relative frequency observer $\cF$\/ has interacted with
all of the $\cO\pup$\/'s, yet there is only a limited sense in which any particular
Everett copy of $\cF$\/ can be said to be ``associated with'' 
a particular copy of one of the $\cO\pup$\/'s. 

There is no  requirement
that such associations exist in the formalism, either from a logical or
physical point of view or in order that it  accord with our
experience.  The perception of a fact regarding the result of
a measurement (or information of any other sort) must be embodied  in the state
of some physical system \cite{Landauer91,Zurek91}; more facts require more systems,
or more states of a given system. In the scenario of  Sec. \ref{sec_FI}
above there exist perceptions of the outcomes of measurements of each of the
$\cS\pup$\/'s as well as perceptions of outcomes of relative-frequency 
measurements. 
To the extent that copies of one observer/measuring device can
be associated with copies of another observer/measuring device or physical system, 
such associations are not necessarily one-to-one.  In the present example,
different copies of
$\cF$\/ may be thought of as ``sharing''  those copies of the $\cO\pup$\/'s 
which are consistent with the
relative frequencies measured by the 
former.  

Rather than a set of ``parallel'' or ``foliated''  universes\cite{Deutsch01}, 
the situation after measurement interactions have taken place has here  a more complicated structure.
E.g., if the perceptions of having measured $\cS^{(1)}$\/,\ldots,$\cS^{(N)}$\/
are stored in respective cell assemblies  $\cO^{(1)}$\/,\ldots,$\cO^{(N)}$\/
in the experimenter's brain, 
and the awareness of the computation of the relative frequency is stored in
a cell assembly $\cF$\/, then the experimenter after having made the measurements
may be  visualized as a ``Siamese $(\nu+1)$\/-tuplet,''  members
of which are distinguished by different copies of $\cF$\/ but share in common copies of the $\cO^{(p)}$\/'s consistent with their differing perceived relative-frequency 
values.

Stein \cite{Stein84}, in criticizing Geroch's 
presentation \cite{Geroch84} of the Everett interpretation, 
comments that ``quantum mechanics will indeed predict with high probability that
the statistics of the outcomes of these experiments [on ensembles of
identical systems] will deviate little from a specifiable set of [relative] frequencies.
But how, in the ordinary practice of physics, do we go about checking this 
prediction? We do so by performing the experiments and {\em noting and
counting their outcomes.''}\/ 
Consistent with this criticism, Farhi, Goldstone and
Gutmann allow that the relative frequency measurements in their analysis 
must be the outputs of  a device specifically constructed to measure relative frequency
{\em without}\/ obtaining any information about the results of measurements of
the individual systems in the ensemble. 
In the present analysis, the individual outcomes
have most certainly been noted, by the explicit interactions with the
$\cO\pup$\/'s, as well as counted by $\cF$\/. 

In the Copenhagen interpretation the weight, say  $W_{f,i}(t_1)$,\/ is of course just the probability 
that the unique outcome of the measurement of $\wh{f}$\/ is $f_i$\/.
The weight  differs from Vaidman's  \cite[Sec. 3.5]{Vaidman02}  ``measure of existence of a world,'' $\mu_i$\/.
The latter is equal to the norm-squared of the   Schr\"{o}dinger-picture branch 
 in which a definite outcome
for the measurement of $\wh{f}$\/  as well as  definite outcomes for measurements of
all the $\wh{a}^{(p)}$\/'s and  $\wh{b}^{(p)}$\/'s  have occurred; i.e., to the
{\em joint}\/ probability for all of these outcomes. 
Vaidman also introduces the concept of a measure of existence for ``I,'' ``the sum of measures
of existence of all different worlds in which I exist...the measure of
existence of my perception world\cite[Sec. 3.5]{Vaidman02}.'' If 
we approximate the totality of my perceptions by my perception of the single
fact of the result of the relative frequency
measurement being $f_i$\/, then $W_{f,i}(t)$\/ is equal to this measure. 
If we enlarge our model
of ``I'' or ``my perception world''  to  include my perceptions of
the measurements of the individual physical systems $\cS\pup$\/, then,
before deciding on an appropriate measure for ``I,'' 
we must address the question of 
how individual perceptions meld together to yield our subjective
sense of unified conscious self-awareness. This is a question,  as-yet-unsolved, 
of neuroscience\cite[p. 464]{KandelSchwartzJessell91}, and as such
involves     neural properties
 which can be described  adequately by
classical physics\cite{Tegmark00}.
As seen above, the lack of an answer to this question 
does not in stand in the way
of constructing locally-realistic quantitative models 
of quantum measurement situations, their outcomes and (using infinite ensembles) the probabilities of these 
outcomes.\footnote{As a candidate for
an operator corresponding to  ``unified conscious self-awareness''
one might consider, following the pattern of  Secs. \ref{sec_MEQM} and \ref{sec_FI},
an additional operator acting on a  $(3^N(\nu+2)+1)$\/-dimensional 
eigenspace with nondegenerate eigenvalues, one of which indicates a
state of ignorance and the remaining $3^N(\nu+2)$\/ of which correspond (via
measurement interactions with the $\cO\pup$\/'s and  $\cF$\/ taking place subsequent to  $U_{\cO}$\/ and $U_{\cF}$)   to the $3^N(\nu+2)$\/ 
distinct sets of   eigenvalues of   the  $\wh{b}\pup$\/'s and $\wh{f}$\/. 
 (This operator could be defined for finite $N$\/ only.) A nonzero weight for a post-measurement
Everett copy of this operator would be equal to the measure of existence of a world,
$\mu_i$\/,
with specific values for the results of measurements of 
$\wh{a}^{(1)}$\/, \ldots, $\wh{a}^{(N)}$\/, and appropriate corresponding
values for the results of measurements of  $\wh{b}^{(1)}$\/, \ldots, 
$\wh{b}^{(N)}$\/ and 
$\wh{f}$\/.}

Regarding the $N \rightarrow \infty$\/ limit: 
The number of post-measurement Everett copies of $\cF$\/ for finite $N$\/ is 
$\nu+1$\/; for infinite $N$\/ it is one or two (see Sec. \ref{subsec_W}).
If we consider the $N$\/ measurements of the $\cO\pup$\/'s to be made sequentially,
with $\cF$\/ recomputing a value of $\phi_i$\/ after each additional $\cO\pup$\/'s
measurement (this would require a different type of interaction than that described
in Sec \ref{sec_FI}), then the weights for the non-Born-rule copies of $\cF\/$\/
decrease with each successive measurement,  and these copies ``finally'' (at $N =\infty$\/) 
vanish altogether from existence.

One can never actually experiment with an infinitely-large 
ensemble,\footnote{This fact does not pose any problems in using the frequency interpretation
of probability as we have  above. Although  we cannot
experiment with an $N \rightarrow \infty$\/ ensemble, this limiting case
is well-defined within the quantum formalism, which, as we have seen,  gives an unambiguous answer as to what would happen if we {\em could}\/ do such experiments.}
so one will never encounter this ``mass extinction'' of Everett 
copies.
Even if it did occur for a physically-realizable size of the ensemble,
it would not be the only situation in which dynamics dictates that
an Everett copy with a certain perception cease to exist.
Vaidman \cite[Sec. 3]{Vaidman98}  discusses a sentient neutron which enters
an interferometer containing at its entrance a beam splitter which splits the trajectory of
an incoming neutron wave-packet into two separate trajectories.
The exit of the interferometer contains another beam splitter which coherently
recombines the two internal trajectories so that the neutron emerges
in a single definite direction.  Immediately before the time at which the neutron
wave packets within the interferometer pass through the second beam splitter,
there are two Everett copies of the neutron, one perceiving ``moving up,''
the other perceiving ``moving down.'' Immediately after the time at which the
wave packets pass through the beam splitter,  only a single Everett
copy remains, perceiving (say) ``moving up.''

The analysis  which I have presented here
introduces
additional structure into the theory beyond the basic Heisenberg-picture formalism; namely,
the notion of existence/nonex\-ist\-ence introduced through the second interpretive rule in
Sec. \ref{subsec_HP}.  
Other frequency-related approaches to probability in the Everett interpretation 
introduce different sorts of additional structure into the theory, such as the  ideas  
of associating a continuous infinity of copies\cite{Deutsch85}
or minds\cite{AlbertLoewer88,Albert92} with each branch of the 
state vector. This  approach can also be applied
to Heisenberg-picture copies \cite{Rubin01}. The number of copies in a branch of the
state vector 
is taken to be proportional to the Born-rule relative frequency, so at any 
time the fraction of the (infinite) total number of copies which perceive any particular
measurement outcome will  be exactly equal to the  probability as given by the
Born rule.  It is an additional assumption to say that this state of affairs,
or, indeed, the corresponding state of affairs in any other approach based on counting the number of outcomes\footnote{Weissman's outcome-counting approach \cite{Weissman99} explicitly modifies the dynamics of the theory, adding
a nonlinear process.}
(branches, copies, minds, \dots),
is  equivalent to the phenomenon of probability. 
Why is it of consequence to me what other copies of me perceive?
In the Everett interpretation in the Heisenberg picture, the number of these continuously-infinite copies or minds will be proportional to the weight. However, if one
introduces the notion of the weight, one can then
consider ensembles of many such measurements, thereby  obtaining
the conclusions obtained in this paper and rendering the {\em ad hoc}\/ introduction of continuous infinities of
copies, minds etc. superfluous.

Zurek \cite[Sec. 4]{Zurek98} uses the fact that the density matrix of a quantum system
subject to environmental decoherence evolves into a form which is for 
all practical purposes identical to the density matrix describing a 
classical statistical ensemble of quantum systems to argue that
the probabilistic interpretation of the latter can be applied to the former.
(See also \cite[Sec. 20]{Gottfried66}.)
Since quantum mechanics is the more
fundamental theory, it seems preferable to derive  probabilistic
behavior purely from within the quantum theory without invoking classical concepts
at the outset. 

Graham's analysis of probability in the Everett interpretation employs two of the
 features which are crucial to the success of the present approach. He 
explicitly includes  the dynamics of a device to measure
relative frequency, and emphasizes that, on physical grounds,
 such a device must have finite resolution.  However, he also introduces the notion
that such a device, as part of a ``two step'' measuring process,\footnote{It may seem that the present approach also involves two
steps, the measurements by the $\cO\pup$\/'s and then the measurement by $\cF$\/.
But we could have considered a finite-resolution relative-frequency
measuring device coupled directly to the $\cS\pup$\/'s and reached the same conclusions. 
The  $\cO\pup$\/'s were included to demonstrate that, as
discussed above, the relative frequency being 
measured is the familiar quantity compatible with measurement of  individual outcomes.}   must first enter a state of thermodynamic equilibrium after it
has made its measurements, and then  must be measured by an additional
observer before 
Everett 
branches emerge in such a way
that the vast
majority of them correspond to perceptions of Born-rule relative
frequencies.    
As shown above, these additional concepts are not necessary
to demonstrate the existence of Born-rule-consistent probability in the 
Heisenberg-picture Everett interpretation.

\section*{Acknowledgments} 

I would like to thank Jian-Bin Mao and Allen J. Tino for helpful
discussions.


\begin{thebibliography}{99}

\bibitem{Everett57}H. Everett~III, `` `Relative state' formulation of quantum 
mechanics, '' {\em Rev. Mod. Phys.}\/ {\bf 29},  454-462 (1957). Reprinted in B.~S.~DeWitt and
N.~Graham, eds., {\em The Many Worlds Interpretation of Quantum Mechanics}
(Princeton University Press, Princeton, NJ, 1973). 

\bibitem{dEspagnat76}B.~d'Espagnat, {\em Conceptual Foundations of Quantum Mechanics}\/,
2nd edn. (W.~A.~Benjamin, Inc., Reading, MA, 1976).

\bibitem{Cramer46}H.~Cram\'{e}r, {\em Mathematical Methods of Statistics}\/
(Princeton University Press, Princeton, NJ, 1946).

\bibitem{Vaidman02}L.~Vaidman, ``Many-worlds interpretation of quantum mechanics,'' in {\em Stanford Encyclopedia of Philosophy}\/ (Summer 2002 Edition), E.~N.~Zata, ed., http://plato.stanford.edu/archives/sum2002/entries/qm-manyworlds.

\bibitem{Polya54}G.~P\'{o}lya, {\em Mathematics and Plausible Reasoning},\/ Vol. II, 2nd. ed.
(Princeton University Press, Princeton, NJ, 1968). 

\bibitem{Hartle68}J.~Hartle, ``Quantum mechanics of individual systems,''
{\em Am. J. Phys.},\/ {\bf 36}, 704-712 (1968).

\bibitem{Finkelstein63}D.~Finkelstein, ``The logic of quantum mechanics,'' 
{\em Trans. N. Y. Aca. Sci.}\/ {\bf 25}, 621-637 (1965).



\bibitem{Graham73}  N.~Graham,    ``The measurement of relative frequency,'' in
B.~S.~DeWitt and N.~Graham, eds., {\em The Many Worlds Interpretation of Quantum Mechanics}
(Princeton University Press, Princeton, NJ, 1973). 

\bibitem{FarhiGoldstoneGutmann89}E.~Farhi, J.~Goldstone, and S.~Gutmann,
``How probability arises in quantum mechanics,'' {\em Ann. Phys. (NY)}\/ {\bf 192},
368-382 (1989).

\bibitem{Kent90} A.~Kent,   ``Against many-worlds interpretations,'' {\em Int. J. 
Mod. Phys.}\/ {\bf A5}, 1745-1762 (1990); gr-qc/9703089.

\bibitem{Zurek98} W.~H.~Zurek,   ``Decoherence, einselection and the existential interpretation (the rough guide),'' {\em Phil. Trans. R. Soc. London}\/ {\bf A 356}, 1793-1820 (1998);
quant-ph/9805065.

\bibitem{Deutsch96} D.~Deutsch,  {\em The Fabric of Reality}\/ (The Penguin Press, 
New York, 1997).

\bibitem{DeWitt72} B.~S.~DeWitt, ``The many-universes interpretation of
quantum mechanics,'' in {\em Proceedings of the International
School of Physics ``Enrico Fermi'' Course IL: Foundations of Quantum Mechanics}\/ (Academic
Press, Inc., New York, 1972). Reprinted in B.~S.~DeWitt and
N.~Graham, eds., {\em The Many Worlds Interpretation of Quantum Mechanics}
(Princeton University Press, Princeton, NJ, 1973). 


\bibitem{Okhuwa93}Y.~Okhuwa, ``Decoherence functional and probability interpretation,''
{\em Phys. Rev. }\/ {\bf D48}, 1781-1784 (1993).

\bibitem{DeutschHayden00}  D.~Deutsch, and P.~Hayden,    ``Information flow in entangled
quantum systems,'' {\em Proc. R. Soc. Lond.}\/ {\bf A456}, 1759-1774 (2000);
quant-ph/9906007.

\bibitem{Rubin01} M.~A.~Rubin,  ``Locality in the Everett interpretation of 
Heisenberg-picture quantum mechanics,'' {\em Found. Phys. Lett.}\/ {\bf 14}, 301-322 (2001);
quant-ph/0103079.

\bibitem{Rubin02} M.~A.~Rubin,  ``Locality in the Everett interpretation of 
quantum field theory,'' {\em Found. Phys.}\/ {\bf 32}, 1495-1523 (2002); quant-ph/0204024.

\bibitem{Vaidman98} L.~Vaidman,   ``On schizophrenic experiences of the neutron or why we should believe in the many-worlds interpretation of quantum theory,'' {\em Int. Stud. Phil. Sci.}\/
{\bf 12}, 245-261 (1998); quant-ph/9609006.

\bibitem{Deutsch99}D.~Deustsch, ``Quantum theory of probability and decisions,''
{\em Proc. R. Soc. Lond.}\/ {\bf A455}, 3129-3137 (1999); quant-ph/9906015.

\bibitem{DeWitt98}B.~DeWitt,  ``The quantum mechanics of isolated systems,''
{\em Int. J. Mod. Phys.}\/ {\bf A13}, 1881-1916 (1998).

\bibitem{Polley99}L.~Polley, ``Quantum-mechanical probability from the symmetries
of two-state systems,'' quant-ph/9906124.

\bibitem{Polley01}L.~Polley, ``Position eigenstates and the statistical axiom of quantum 
mechanics,'' quant-ph/0102113.

\bibitem{CavesFuchsSchack}C.~M.~Caves, C.A.~Fuchs, and R.~Schack,
``Quantum probabilities as Bayesian probabilities,''  {\em Physical Review}\/
{\bf A65}, 022305 (2002); quant-ph/0106133.

\bibitem{Beckenstein01} J.~D.~Beckenstein, ``Limitations on quantum information
from black hole physics,'' {\em Acta Phys. Polon.}\/ {\bf B32}, 3555-3570 
(2001); quant-ph/0110005.

\bibitem{Renyi70} A.~R\'{e}nyi, {\em Foundations of Probability}\/ 
(Holden-Day, Inc., San Francisco, 1970).

\bibitem{Rudin76}W.~Rudin, {\em Principles of Mathematical Analysis},\/ 3d edn. 
(McGraw-Hill, New York, 1976).

\bibitem{Griffiths84}R.~B.~Griffiths, ``Consistent histories and the
interpretation of quantum mechanics,'' {\em J. Stat. Phys.}\/,
{\bf 36}, 219-272 (1984).

\bibitem{Hartle95}J.~B.~Hartle, ``Spacetime quantum mechanics and the quantum mechanics 
of spacetime,'' in B.~Julia and J.~Zinn-Justin, eds., {\em Les Houches, Session LVII, 1992, Gravitation and Quantizations}\/ (Elsevier Science B.V., 1995).

\bibitem{Page94}D.~N.~Page,``Probabilities don't matter,'' in M.~Keiser and
R.~T.~Jantsen, eds., {\em Proceedings of the 7th Marcel Grossmann
Meeting on General Relativity},\/ (World Scientific, Singapore, 1995); gr-qc/9411004.

\bibitem{Page95}D.~N.~Page,``Sensible quantum mechanics: Are only perceptions probabilistic?,''
quant-ph/9506010.

\bibitem{Page96}D.~N.~Page, ``Sensible quantum mechanics: Are probabilities 
only in the mind?,'' {\em Int. J. Mod. Phys.}\/ {\bf D5}, 583-596 (1996); gr-qc/9507042.


\bibitem{Landauer91}R.~Landauer, ``Information is physical,'' {\em Physics Today} {\bf 44},
May, 23-29 (1991).

\bibitem{Zurek91}W.~H.~Zurek, ``Decoherence and the transition from quantum to 
classical,'' {\em Physics Today}\/  {\bf 44}, Oct., 36-44 (1991).

\bibitem{Deutsch01}D.~Deutsch, ``The structure of the multiverse,'' {\em Proc. R. Soc. Lond.}\/ {\bf A458}, 2911-2923 (2002); quant-ph/0104033.

\bibitem{Stein84} H.~Stein, ``The Everett interpretation of quantum mechanics: Many worlds
or none?,'' {\em No\^{u}s}\/ {\bf 18}, 635-652 (1984).

\bibitem{Geroch84} R.~Geroch, ``The Everett interpretation,'' {\em No\^{u}s}\/ {\bf 18}, 617-633 (1984).

\bibitem{KandelSchwartzJessell91}E.~R.~Kandel, J.~H.~Schwartz, and T.~M.~Jessell,
{\em Principles of Neural Science},\/ 3d. edn. (Appelton \& Lange, Norwalk, CT, 1991).

\bibitem{Tegmark00}M.~Tegmark, ``The importance of quantum decoherence in brain processes,''
{\em Phys. Rev.}\/ {\bf E61}, 4194-4206 (2000); quant-ph/9907009.

\bibitem{Deutsch85}  D.~Deutsch,``Quantum theory as a universal physical
theory,'' {\em Int. J. Theor. Phys.}\/  {\bf 24}, 1-41 (1985).

\bibitem{AlbertLoewer88}  D.~Albert and B.~Loewer,   ``Interpreting the many worlds 
interpretation,'' {\em Synthese}\/ {\bf 77}, 195-213 (1988).

\bibitem{Albert92}  D.~Z.~Albert,   {\em Quantum Mechanics and Experience}\/ (Harvard University
Press, Cambridge, MA, 1992).

\bibitem{Weissman99} M.~B.~Weissman,  ``Emergent measure-dependent probabilities from modified quantum dynamics without state-vector reduction,'' {\em Foundations of Physics Letters}\/ 
{\bf12}, 407-426 (1999); quant-ph/9906127.

\bibitem{Gottfried66} K.~Gottfried, {\em Quantum Mechanics, Vol. I: Fundamentals}\/ (W.~A.~Benjamin, Inc., Reading, MA,
1966).

\end{thebibliography}
\end{document}